\documentclass[prl,twocolumn,superscriptaddress,showpacs]{revtex4}
\usepackage{graphicx}
\usepackage{amssymb,amsmath,amsfonts}
\usepackage{wasysym}
\usepackage{latexsym}

\newcommand{\be}{\begin{eqnarray}}
\newcommand{\ee}{\end{eqnarray}}

\begin{document}
\title{Surprise(s) in magnets without net moments}
\author{Kedar Damle}
\affiliation{\small{Tata Institute of Fundamental Research, 1, Homi Bhabha Road, Mumbai
400005, India}}
\date{January 12, 2009}

\begin{abstract}
We are all familiar with ferromagnetic and antiferromagnetic materials, in which
the localized ionic moments (in case of ionic insulators)  or the electronic spins
(in case of metals)
go into a long-range ordered state with a net macroscopic moment (in case
of ferromagnets) or a net macroscopic sublattice magnetization (in case of antiferromagnets).
However this behaviour is far from ubiquitous even in ionic insulators with well-developed
local moments. Indeed, there are many ionic insulators in which the
dominant interactions between
the local moments compete with each other, leading to a cooperative paramagnetic
state with no ordering of the moments down to the lowest temperatures accessible to
experiments. The physics of such magnets without net moments has some interesting
and surprising aspects, which are touched upon in this brief review.
\end{abstract}

\vskip2pc

\maketitle

{\em Overview:}
In many ionic insulators, some of the ions have non-zero ground-state
angular momemtum, and an associated magnetic moment. These localized
magnetic moments usually interact with each other in two ways: On
the one hand, by virtue of a being magnetic dipoles, each moment reacts to the dipole field of all other magnetic
dipoles, giving rise to a long-range anisotropic magnetic dipolar interaction.
On the other hand, the virtual hopping of charge carriers between
neighbouring magnetic ions gives rise to a short-ranged (typically nearest
neighbour) interaction generally known as the exchange
interaction~\cite{Goodenough}.

This exchange energy can be written as
$E =  J \sum_{\langle i j \rangle} {\mathbf S}_i \cdot {\mathbf S}_j  \; ;J>0 $,
where $J$ is the exchange constant and the subscripts refer to pairs of nearest-neighbour moments. In
many commonly occuring cases, $J$ is positive and therefore antiferromagnetic in
nature~\cite{Goodenough}, in that
it encourages antiparallel alignment of pairs of neighbouring moments. In many
situations, this exchange constant is much larger than the long-range
dipolar coupling, which can then be left out of the analysis
to a very good approximation, while in other cases, the dipolar
coupling can dominate over the nearest-neighbour exchange.
[Here, the spins
${\mathbf S}$ are of course quantum-mechanical operators; however, for many purposes at not too low temperatures,
they can be usefully approximated by classical vectors of fixed length, particularly if the spin quantum number
$S$ is $3/2$ or higher.]

When the magnetic ions form a bipartite lattice (in which the lattice can be broken up into
two sublattices in such a way that the nearest neighbour exchange coupling connects
only pairs of spins belonging to different sublattices), this antiferromagnetic exchange energy is minimized by the
so-called {\it Neel} state in which all spins
lie along a spontaneously chosen axis ${\bf n}$ and every spin points anti-parallel to its nearest neighbours
[In two and higher dimensions, this picture also gives an essentially correct caricature of the
ground state of the full quantum problem on a square or hypercubic lattice.] This is the case, for
instance, in the compound MnO which has a Neel ordering temperature of
approximately 116 K.

\begin{figure}
\begin{center}
 \includegraphics[width=3cm]{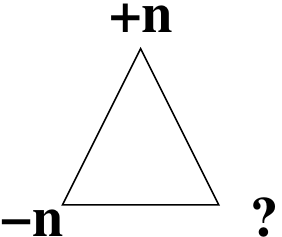}
 \caption{Three spins interacting antiferromagnetically with each other cannot satisfy the demands of all the exchange interactions}
\end{center}
 \label{triangles}
 \end{figure}

However, there are many other examples in which the magnetic ions are coupled by exchange
couplings that do not obey this bipartite constraint---in other words, there are triangles in the
nearest neighbour connectivity of the lattice.
Such magnetic lattices with triangular motifs
in them are good examples of {\em geometric frustration}. To see the significance of
such triangular motifs, it is enough to note that the Neel (antiferromagnetic) state along any axis ${\bf n}$  is {\it frustrated} in the presence of such triangles, since there is no unique
way of satisfying all the exchange interactions (Fig~\ref{triangles}) {\em fully}. 

In many situations~\cite{exptreview1}, this results in a macroscopic degeneracy of {\it classical} minimum energy configurations.
At intermediate temperatures $T$ that are less than the exchange $J$, but are not small enough for the quantum mechanical nature of spins to matter, the spin correlations (measured,
say, by neutron scattering experiments) in the system simply reflect this macroscopic degeneracy, and can be modeled in a universal
way in terms of averages over an ensemble that gives a certain weight to each of these minimum energy configurations~\cite{Moessnerreview}. To a first approximation, this weight is of course uniform and the same
for each minimum energy microstate, and the sub-leading effects of canonical fluctuations (that increase
the energy from this minimal value) can also be included in a more sophisticated treatment. 
Many examples of such frustrated magnets are known. On the pyrochlore lattice, these
include the $Cu^{2+}$ based $S=1/2$ magnet paramelaconite~\cite{pyro1} and the $Cr^{3+}$ based $S=3/2$ magnets
CdCr$_2$O$_4$ and HgCr$_2$O$_4$~\cite{pyro2}. Several interesting examples have also been studied
on the kagome lattice---these include $Cu^{2+}$ based $S=1/2$
volborthite and other systems\cite{otherkagome0}, $Ni^{2+}$ based $S=1$ magnets Ni$_3$V$_2$O$_8$\cite{nickel1}, $Cr^{3+}$ based $S=3/2$ systems
\cite{otherkagome1}, and $Fe^{3+}$ based $S=5/2$ magnets Fe jarosite
\cite{otherkagome2}.

Much of the interest in frustrated magnetism arises from the non-trivial nature
of the resulting state. In particular, as we will see explicitly in examples below,
the resulting intermediate state is not a simple paramagnet with each moment fluctuating more
or less independently of the others. Instead, it is typically a non-trivial cooperative
paramagnet, with a complicated pattern of correlations between far away moments. Such
a cooperative paramagnetic state can have some very unusual properties, and that is
the really interesting thing about these systems.
Finally, it is also useful to keep in mind that the ultimate fate of such magnets at very low temperatures 
is less universal, and depends sensitively on the
effects of quantum fluctuations and other (subdominant) interactions  acting in this subspace.

{\em Toy model---The triangular lattice antiferromagnet:}
Fortunately for us, the triangular lattice Ising antiferromagnet provides a simple
example where a great deal of the foregoing can be illustrated quite explicitly. Here `Ising' refers
to the fact that we consider a simplified situation in which `spins' that are not unit vectors,
but instead discrete variables that can take two values $\pm 1$. Such models are generally referred
to as Ising models, after Ising, who studied them first in his Ph.D thesis.
Although this model looks very over-simplified at first sight, it does have
the potential to describe real magnets at least in some cases.
This is because real magnets have local moments that
often correspond to ionic ground state multiplets with a non-zero value of
orbtital angular-momentum. In such cases, spin-orbit coupling in the presence
of strong crystal field effects can induce a {\em single-ion anisotropy} term 
$-D\sum_i (S^z_i)^2$ in the magnetic Hamiltonian. If $D$ is large (compared to $J$) and
positive, then the spins predominantly prefer to be in one of two states $S^z = \pm S$
(for spin $S$ moments), which can be thought of as the two Ising states in our foregoing
description [One example of this is the Kagome lattice antiferromagnet Nd-Langasite,
where a description in terms of a Ising magnet on the Kagome lattice apparently works quite well]

With that background, we now ask: What configurations minimize the nearest neighbour Ising exchange energy
$E = J \sum_{\langle i j \rangle} \sigma_i \sigma_j$? Clearly, the answer is all configurations in
which each triangle has either two `up' ($+1$) spins and one `down' ($-1$) spin, or vice versa.
These configurations minimize the `frustration' induced by the competing antiferromagnetic
interactions by ensuring that each triangle has exactly one frustrated bond (pair of parallel Ising
spins).

Furthermore, these minimally frustrated configurations have a relatively `clean' characterization
in terms of dimer covers of the dual honeycomb lattice. More explicitly, consider the honeycomb
net formed by forming links between the centers of triangles across the shared side of the triangle.
If we place hard-rods on each honeycomb link that crosses a frustrated bond in a minimally
frustrated configuration, then each honeycomb lattice site will have exactly one hard-rod covering
it. Such configurations of hard-rods define so-called `dimer covers' of the honeycomb lattice,
and clearly, there is a one-to-one mapping between dimer covers of the honeycomb
lattice and minimum frustration states of the classical Ising model on the triangular lattice.

As the temperature $T$ falls well below the exchange energy scale $J$, most triangles of
the lattice will satisfy the minimum frustration condition and have exactly one frustrated bond.
Indeed, one expects that a typically low-temperature configuration will differ
from a minimally frustrated $T=0$ configuration only by an exponentially small (${\mathcal O}(e^{-J/k_BT})$) density of triangles with {\em three} frustrated bonds. If we ignore the effects of such defects, the $T \ll J$
properties of this system can thus be modeled by calculating the properties of
the ensemble of minimum frustration states, with equal weight to each such state.

This is
most conveniently done in dimer language, and one learns two important things upon
translating back to the language of Ising spins: The first is that the Ising spins are
not ordered even at $T=0$, i.e there will be no magnetic Bragg peaks even in a hypothetical neutron
scattering experiment performed on our Ising magnet. The second is that the correlations between
spins do not decay away to zero on the scale of a few lattice spacings, as would
be expected for a simple paramagnet in which the moments fluctuate independently of each
other. Instead, the correlations decay to zero very
slowly, as the inverse square-root of the separation between the spins.

The $T=0$ state
is thus a {\em cooperative paramagnet} in the moments are correlated
with each other over macroscopic distances although there is no long-range ordering.
In this simple toy model, it is also possible to put back the exponentially small density
of defect triangles into our description, and it can be shown that these do not
change the $T=0$ picture in any striking way---all that happens is that the defect
triangles disrupt the slow power-law decay of correlations beyond a length-scale $\xi$ that corresponds
to the typical inter-defect distance, and the correlations decay exponentially
rapidly to zero for $r \gg \xi$.

There are thus two different but related questions that one needs to keep in mind
when thinking about the low temperature properties of such magnets: The first is
the nature of the degenerate minimum exchange energy configurations, and the ensemble
they define. In particular,
in the limiting case of zero temperature, physical quantities can be modeled by
averages over this restricted ensemble of minimum energy configurations. The second
is the nature of the thermally induced defects that allow a system to locally deviate
from a minimum energy configuration, and the properties of a dilute, extremely cold gas of these
defects---the low temperature properties of the magnet depend both on the nature
of the ground state ensemble, and the statistical mechanics of the defect gas.

\begin{figure}
{\includegraphics[width=\hsize]{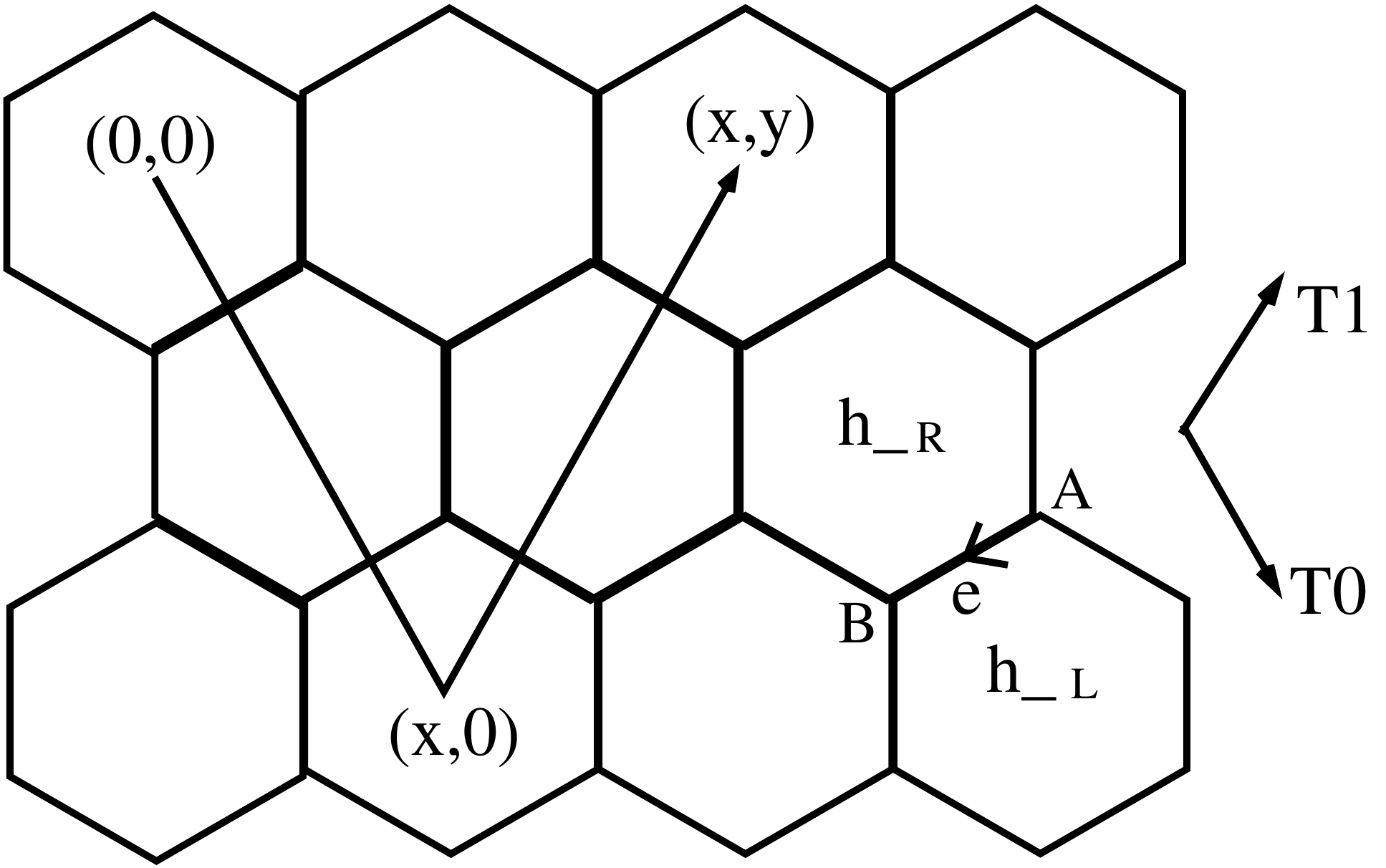}}
\caption{The honeycomb lattice dual to the triangular lattice, the definition of
the electric field and heights, and the path followed to reach point $x,y)$ on the
triangular lattice starting from the origin $(0,0)$}
\label{honeycomb_dimer_height}
\end{figure}

{\em Effective field theory for the $T \rightarrow 0$ limit:}
In order to prepare the ground for our later discussion, it is useful to spend a little
time understanding these results from the perspective of a coarse-grained effective
field theory. The idea is think of the dimer occupation on a link as the value of
an electric field ${\bf e}$ on the corresponding link of the honeycomb lattice, with
the sign conveniton that the ${\bf e}$ on each link always points from the $A$ sublattice
site to the $B$ sublattice site of that link (see Fig~\ref{honeycomb_dimer_height}).
Clearly, the dimer constraint now translates to the statement that there is a static $+$ charge
on each $A$ sublattice site of the honeycomb net, and a static $-$ charge on each $B$ sublattice
site. Now, we solve for this Gauss's law divergence constraint by writing ${\bf e}$
in terms of a {\em height field} (which is the two dimensional analog of the vector potential
of ordinary electrodynamics).

This height field $h$ is defined on the original triangular lattice sites,
but is quite distinct from the original spins, and in terms of $h$, we may write the electric
field on link $l$ as
\begin{equation}
e_{l} - \frac{1}{3} = h_{L(l)} - h_{R(l)}
\end{equation}
where $R(l)$ and $L(l)$ are triangluar lattice sites to the right and left of this link (as defined
when looking down the link from its $A$ sublattice end (see Fig~\ref{honeycomb_dimer_height}).

We now note that the {\em staggered} dimer configuration in which we occupy all links of one orientation
and leave all others free corresponds to a height configuration with maximum {\em tilt} (gradient) in one
of the  principal
directions of the triangular lattice. By inspection, we also see that such a staggered configuration cannot be
changed into
any other dimer configuration by any local moves that do not involve a macroscopically large number of
links, and thus has {\em very few nearby} dimer configurations. Conversely, dimer configurations
that can be transformed into other valid dimer configurations in a large number of ways tend
to have zero average tilt in height language.

With this motivation, one postulates a coarse-grained
`free energy' that captures the entropic weight of different height configurations
and writes the $T=0$ partition function as
\begin{equation}
Z = \int {\cal D} h(x) \exp \left (-\frac{K}{2}\int d^2 x (\nabla h)^2 \right ) \, ,
\end{equation}
where $K$ is a phenomenological `stiffness' parameter.
Of course, in order to use this effective field theory, one needs a prescription for writing
the local spin density $\sigma({\mathbf r})$ in terms of the height fields. To understand this
correspondence, one may start by fixing one spin, say the spin on the site at the origin, to be up
$\sigma({\mathbf r}=0) = +1$, and the corresponding height to be zero $h({\mathbf r} = 0) = 0$. Now,
we note that $3h$ jumps by an odd number whenever one crosses an unfrustrated bond (across which
the spin flips sign), while the height jump is even across a frustrated bond (across which
the spin remains unchanged).

This immediately implies that
a spin at some other site ${\mathbf r}$ 
will be up if and only if $3h({\mathbf r})$ is even~\cite{Nienhuis_hilhorst_blote}. This provides one piece of the correspondence
between the height field and the spin field. The second, and in many contexts more crucial, piece
of the correspondence is slightly trickier to understand, and is best appreciated
by trying to predict the spin value at site ${\mathbf r} = (x,y)$ by going across $x$ links of
the dual honeycomb lattice in direction $T_0$, followed by $y$ links of the dual
lattice in direction $T_1$~\cite{Heidarian_damle_unpublished}. Now, each dimer crossed in the process guarantees that the spin
state has not changed, while each empty link corresponds to a flip in the spin state.
We may therefore write $\sigma(r) = \exp\left(i\pi \sum_l (1-n_l)\right) \sigma(0)$, where
$n_l$ is the dimer number on all the links $l$ thus encountered. Rewriting this in
terms of the height field allows us to argue that
$\sigma({\bf r}) \sim \exp\left(2\pi i(x+y)/3 + i \pi h(x,y)\right) + h.c$.

Thus, both the zero momentum and momentum $\pm Q \equiv \pm (2\pi/3,2\pi/3)$ components of the
spin field have a simple local representation in terms of exponentials of the height field,
and the long-distance properties of the spin correlations may be captured
by the correspondence: $\sigma({\bf r})= c_{Q}\cos\left(\pi h ({\bf r}) + 2 \pi (x+y)/3\right)
+ c_0\cos(3\pi h({\bf r}))$, where $c_Q$ and $c_0$ are non-universal scale factors. Using this and the known value of $K$~\cite{Nienhuis_hilhorst_blote}, we can calculate the $T=0$ correlators of
the Ising spins, and find that the leading term at large separation $r$
goes as $\cos(2\pi r/3)/\sqrt{r}$. Thus, the triangular Ising magnet is
anything but a simple uncorrelated paramagnet, although it does not order
at all even at $T=0$---instead, it forms a non-trivial correlated
state with very slowly decaying correlations.

This simple example illustrates the non-trivial nature
of the cooperative paramagnetic state of frustrated magnets.
Such non-trivial correlated states can have interesting properties, including
unusual low-lying excitations. Below, we will describe one 
example of this in some detail, and briefly allude to another.

\begin{figure}

{\includegraphics[width=\hsize]{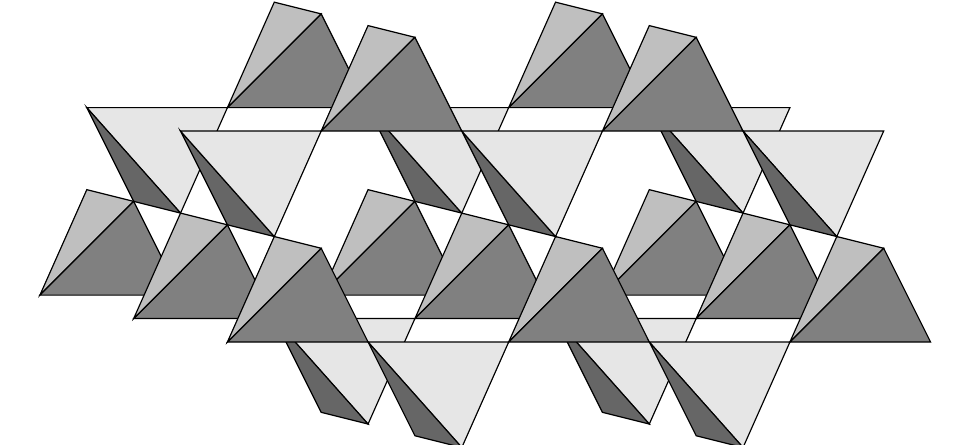}}
\caption{The pyrochlore structure made up of corner sharing tetrahedra whose centers form the diamond lattice. In the spin-ice compounds, Ho$^{3+}$ or  Dy$^{3+}$
moments occupy vertices of the tetrahedra shown forming spin-ice.}
\label{example1}
\end{figure}

\begin{figure}

{\includegraphics[width=\hsize]{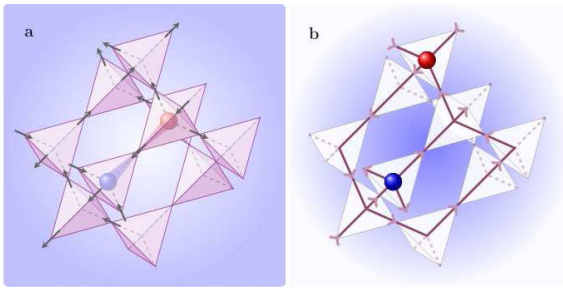}}
\caption{Tetrahedra violating the two-in two-out ice-rule are actually
magnetic monopoles. If they were bound in nearest-neighbour pairs (as on left), then
the lowest lying excitations would be flipped spins, but they are
in fact unbound (as on right), interacting only through a magnetic analog of Coulomb's law!
Note the reversal of the colour convention on the right (Figure taken
from the arXiv version of Ref.~{\protect{\onlinecite{Jaubert_holdsworth_naturephysics}}})}
\label{dumbells}
\end{figure}

{\em Magnetic monopoles in spin-ice materials:}
With this background, we now come to the first surprise we advertised
in our title and abstract, namely, the existence of genuine
{\em magnetic monopoles} in the low-energy spectrum of the
easy-axis pyrochlore lattice antiferromagnets Dy$_2$TiO$_3$ and Ho$_2$TiO$_3$, that are generally
refered to as {\em spin-ice} compounds
[The interested reader should
also refer to the original articles Refs~\onlinecite{Castelnovo_moessner_sondhi_nature,Jaubert_holdsworth_naturephysics,Isakov_moessner_sondhi_prl}
for a more detailed discussion of the various technical aspects and subtleties involved.]

In these materials,  Dy$^{3+}$ and Ho$^{3+}$ occupy the vertices of the pyrochlore lattice
tetrahedra (shown in Fig~\ref{example1}) and  carry a ground state magnetic dipole moment
$\mu = 10\mu_B$ (where $\mu_B$ is the Bohr magneton) that has its origins in
the spin-orbit coupled ground state multiplet for this valence state. Crystal field
effects result in a strong `easy-axis' energy that forces eacn moment
to lie along the tetrahedral body diagonal that passes through the corresponding
pyrochlore lattice site. Thus, if one considers a single tetrahedron, each moment
has two choices: It can either point inward towards the center of the tetrahedron, or
outward away from the center of the tetrahedron---in both cases, it must lie
precisely along the corresponding body-diagonal.

This degree of freedom can be thought of as an Ising spin $\sigma$, and the
magnetic properties of these materials can again be modeled as some sort
of Ising model. Since each dipole is shared by one up-pointing tetrahedron
and one down-pointing tetrahedron, a convenient sign-convention for
this mapping is that $\sigma=+1$ if the corresponding dipole points outwards
when viewed from the up-pointing tetrahedron to which it belongs. Conversely,
$\sigma=-1$ if the corresponding dipole points outwards
when viewed from the {\em down-pointing} tetrahedron to which it belongs.

What is the Hamiltonian or energy functional that describes the energetics
of these Ising spins? The answer is a little complicated: It turns out that
the nearest neighbour exchange coupling in these systems is weak (of order $1K$ in
temperature units), and the long
range magnetic dipole interactions between the magnetic moments is actually
more important.

The Hamiltonian is thus the well-known classical expression for the interaction
energy of a bunch of magnetic dipoles, oriented along body-diagonals of the tetrahedra.
Rather than write this big expression down explicitly, it is useful to use a pedagogical
device and think in terms of a `dumbell-model'~\cite{Isakov_moessner_sondhi_prl} for the interaction energy.
The idea is quite simple: We know that the interaction energy between two spatially separated
groups of
electric charges, each of which is overall charge-neutral, can be approximated by a multipole
expansion, of which the dipole-dipole interaction energy is the leading term at large distances,
with corrections that fall off as a faster power of the distance between the two
groups of charges. We can turn this standard fact around, and view each magnetic
dipole as being made up of a dumbell with a `blue' and `red' end located at
the body-centers of the two tetrahedra that share the pyrochlore lattice site
on which the magnetic dipole is located (as shown in Fig~\ref{dumbells}). These blue and red ends thus lie
on sites of the dual diamond lattice whose sites are the body-centers of
the pyrochlore tetrahedra, and whose links pass through the sites of the pyrochlore
lattice.

The blue end represents a
fictitious positive magnetic charge $+q_m/2$, and the red end represents a fictitious negative
magnetic charge $-q_m/2$ (the reason for the factor of two in this definition will be clear below). The magnitudes of these charges are adjusted to give the correct
magnetic dipole strength of $\mu = 10\mu_B$ by requiring that $q_m/2 = \mu/a_d$, where $a_d$
is twice the distance from the vertex of a tetrahedron to its body-center along
the body diagonal (equivalently, $a_d$ is the nearest-neighbour distance of the dual
diamond lattice). The Ising degree of freedom $\sigma$ at each pyrochlore
site now corresponds to the orientation of this dumbell, and the Ising spin
$\sigma$ at the vertex of an up-pointing tetrahedron is $+1$ if
the red end of the dumbbell is located at its body-center, and
$-1$ if the blue end of the dumbbell is located at its body-center.

The original energy functional
can now be simply (but approximately) reproduced by postulating a fictitious Coulomb interaction
between these fictitious magnetic charges:
\begin{eqnarray}
V_m(r_{\alpha \beta})  & = &\frac{\mu_0}{4\pi} \frac{Q_\alpha Q_\beta}{r_{\alpha \beta}}\; ; \alpha \neq \beta  \nonumber \\
&& = \frac{1}{2}v_0Q_\alpha^2 \; ; \alpha=\beta \nonumber
\end{eqnarray}
Here $\mu_0$ is the vacuum permeability, $Q_\alpha$ is the total magnetic charge on
diamond lattice site $\alpha$ (corresponding to the body-center of tetrahedron $\alpha$)
and the `self-energy' constant $v_0$ is adjusted to correctly reproduce the interaction
energy between nearest neighbour dipoles.

Naturally, this statement is only approximate, but the approximation involved is such that the difference
between the real interaction energy and the approximate form obtained
by our device of introducing fictitious magnetic charges falls off rapidly with distance
$r$ between the magnetic dipoles, and is very small everywhere.

This way of thinking immediately yields dividends when we ask for the nature of
the minimum energy configurations of the Ising spins $\sigma$. Using
the electrostatic analogy, it becomes clear that the minimum energy configurations
are precisely those configurations for which the total (fictitious) magnetic charge on
the body-center of {\em each tetrahedron} is zero $Q_\alpha = 0$ for all $\alpha$.
Translated to Ising variables, this is the `two-in two-out' `ice rule' that
says that two vertices of each tetrahedron must have Ising spins pointing inwards
towards its body-center, while two must have Ising spins pointing outwards
away from its body-center. [Here `ice-rule' refers to the analogy to Pauling's ideas
about the entropy of ice, and Nagle's unit model for ice~\cite{Nagle}].

How many such minimum energy configurations are there? The answer is that
the set of minimum energy configurations has macroscopic entropy, {\em i.e} it scales
as the exponential of the number of system sites. This is closely analogous to the
triangular lattice example we discussed earlier, and indeed, the spin-ice
minimum energy configurations can also be characterized in terms of dimer configurations
as before: One simply considers all up-pointing tetrahedra, and agrees to
put a dimer through each outward pointing magnetic moment. To ensure consistency,
the rule is reversed for all down-pointing tetrahedra, that is, each
inward pointing spin of a down-pointing tetrahedron corresponds to a dimer on
the diamond lattice link passing through that spin. This gives a dimer model
on the diamond lattice, with two dimers touching each diamond lattice vertex,
and one can then use efficient {\em loop algorithms} to sample all the
minimum energy configurations in this dimer representation~\cite{Sandvik_moessner,Banerjee_damle_unpublished}.

What about excited states? Naively, one might imagine that the lowest lying
excited states may be constructed by starting with an arbitrary
minimum energy configuration, and flipping one Ising spin. Such a flipped
spin would give rise to two nearest neighbour tetrahedra that
violate the ice rule---one of them will have three out-pointing spins,
and one of them will have three in-pointing spins.

However, the language
of fictitious magnetic charges allows us to think a little bit more
deeply. For consider flipping a single spin $\sigma_{\alpha \beta}$
from $+1$ to $-1$.
This creates two equal
and opposite fictitious magnetic charges at nearest neighbour sites $\alpha$ and $\beta$:
$Q_\alpha = +q_m$, $Q_\beta = -q_m$. Thought of in this way, there is nothing
special about having these two charges $\pm q_m$ at nearest neighbour locations.
By flipping a string of Ising spins, we can pull these charges further and further apart
from each other until they are separated by distance $r$. Now, these two charges
interact with an attractive Coulombic potential that falls off as $1/r$.
As we know, a $1/r$ attraction is not confining, in the sense that it only
takes a finite amount of work to separate the two charges to infinity.
Thus, the dominant excitations will {\em not} correspond to two charges
$\pm q_m$ bound tightly at nearest-neighbour distance, but rather two
independent and `free' charges that can be at arbitrary separations from
each other.

We may now translate back to the language of the original magnetic moments and
Ising spins: The dominant low energy excitations consist of arbitrarily long
strings of flipped Ising spins. Some more analysis reveals that the end points of these strings are
genuine
magnetic monopoles of strength $\pm q_m$, in the sense that the associated magnetic field configuration
would induce a current in a superconducting ring if one end of the string
passed through the ring. This current would be identified
 as a monopole signal in a standard monopole search experiment
such as the Stanford experiment to detect fundamental magnetic monopoles in
cosmic radiation!

So far, all of this is strictly a $T=0$ argument about ground states and
low-lying excited states. One may worry that entropic effects associated with
thermal fluctuations at non-zero temperature would somehow spoil all this.
Perhaps fortuitously, the answer is no! It turns out that the most
important consequence of entropic effects is a modification of the prefactor
of the Coulombic attraction between our emergent magnetic monopoles, by an
amount proportional to $T$ that can be calculated precisely by computer
simulations~\cite{Huse_etal,Henley_pyrochlore,Banerjee_damle_unpublished}. This is a fairly innocuous effect, and is
not expected to change the basic picture outlined above.
\begin{figure}

{\includegraphics[width=\hsize]{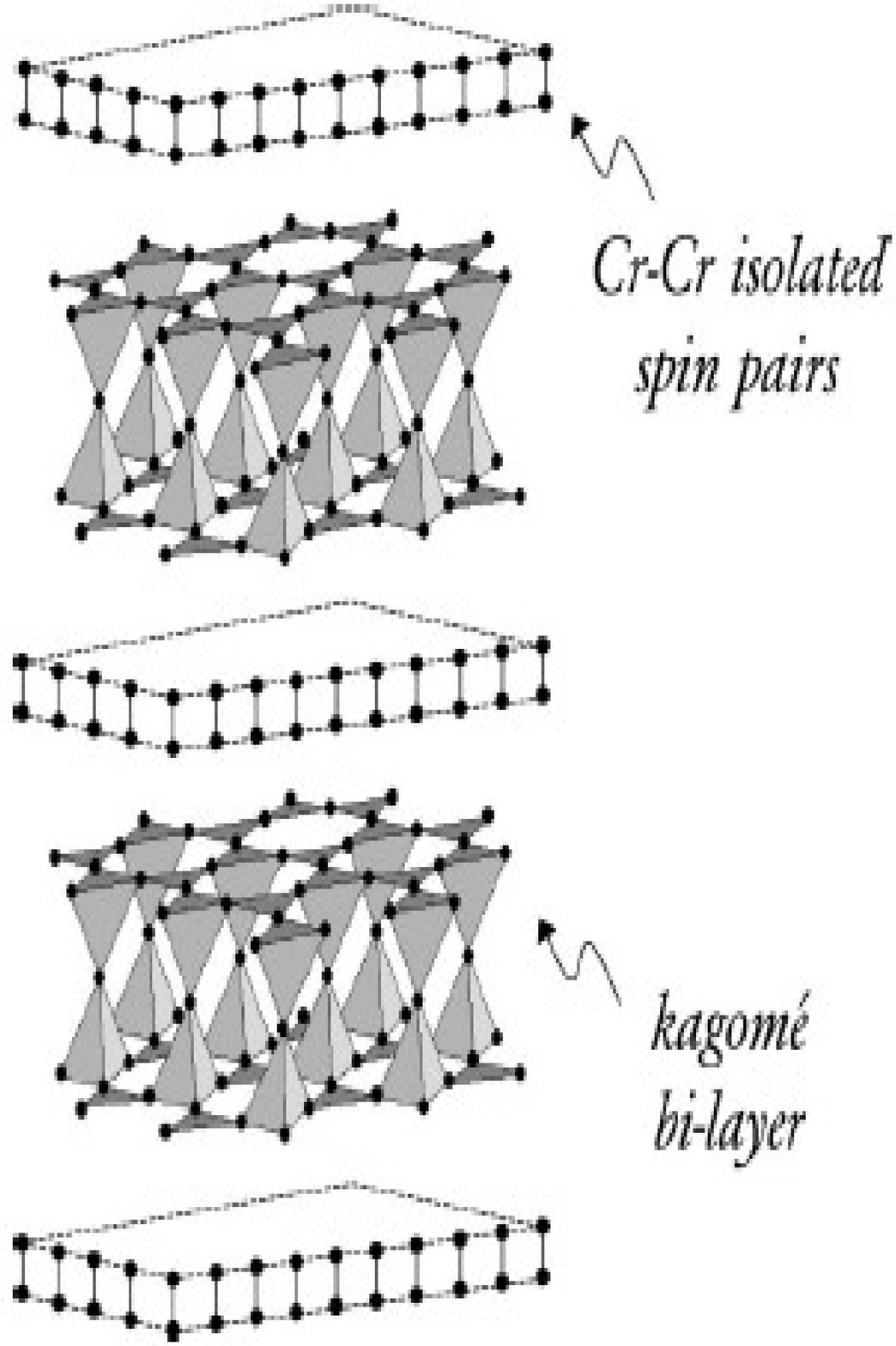}}
\caption{The kagome bilayer structure of SCGO, with each Kagome layer made up of corner sharing triangles, and joined to the other by apical sites. Cr$^{3+}$ ions with $S=3/2$ occupy the corners of these triangles as well as the apical sites. Other Cr$^{3+}$ ions
form the spacer layers consisting of isolated spin dimers  (Figure
is taken from the arXiv version of Ref~{\protect \cite{Limot_etal_prb}})}
\label{example2}
\end{figure}

{\em Impurity induced half-orphan $S=3/4$ spins in SCGO:}
The second of our promised surprises is Henley's~\cite{Henley_2000} identification
of the unusual defect induced local moments that may exist in
the pyrochlore slab magnet SCGO. SCGO is an abbreviation
for the oxide SrCr$_9$Ga$_3$O$_{19}$, where the listed
formula is only notional since this ideal stoichiometric composition
can never be in the laboratory. What is instead
commonly prepared is  SrCr$_{9p}$Ga$_{12-9p}$O$_{19}$, with $p$ ranging
all the way from roughly $0.5$ to $0.98$~\cite{Limot_etal_prb}.

The ideal stoichiometry corresponds
to $S=3/2$ Cr$^{3+}$ ions occupying the sites of a `pyrochlore slab', in addition
to forming a layer of decoupled pairs (Fig~\ref{example2}). The spins in the pyrochlore slab interact with
each other through an isotropic nearest neighbour antiferromagnetic exchange coupling $J \approx 100K$ (in temperature units), forming two
Kagome layers coupled through apical sites in the middle (Fig~\ref{example2}).
The spins in the dimer layer only interact within a pair via an isotropic antiferromagnetic
exchange $J {'} \approx 200K $ (in temperature units), thus forming
a system of decoupled spin-dimers.
The excess Ga introduced by having $p < 1$, substitute for the Cr$^{3+}$ ions
and introduce non-magnetic impurities with $S=0$. From detailed studies~\cite{Limot_etal_prb},
it is known that the Ga have a slight preference for going into
the Kagome and isolated dimer layers, rather than substitute for the apical
Cr$^{3+}$. 

With this background, let us now think classically ($S=3/2$ is large enough
that we expect a classical analysis to be accurate except at extremely
low temperatures at which quantum fluctuations start to play an important role)
and ask for the classical minimum energy configurations of the exchange
Hamiltonian
\be
 H = \frac{J}{2}\sum_{\XBox} (\sum_{i \in \XBox}\vec{S}_i - \frac{\mathbf{h}}{2J})^2 + \frac{J}{2}\sum_{\triangle} (\sum_{i \in \triangle}\vec{S}_i - \frac{\mathbf{h}}{2J})^2 \nonumber
\label{eq2}
\ee
where $\XBox$ refers to the tetrahedra that have as one of
their faces the up-pointing (down-pointing) triangles in the upper (lower)
Kagome layer, and $\triangle$ refers to the down-pointing (up-pointing)
triangles in the upper (lower) Kagome layer, and ${\mathbf h}$ is the external
magnetic field, which we now proceed to set to zero.

When written in this form, it is clear that this energy functional has enormously
many ground states, which correspond to all possible configurations
in which each simplex (a tetrahedron or a triangle) has zero net spin.
Let us now dope the system with non-magnetic impurities.
If a simplex has a single non-magnetic site, it can still arrange the
spins on the other sites to add up to zero, and thus a single vacancy
on a simplex has no significant effect on the properties of the system.

What about a correlated defect consisting of two vacancies on a single simplex?
If this simplex is a tetrahedron, again, nothing much happens. However, if this
simplex is a triangle, then it becomes impossible to satisfy the
zero spin constraint on this simplex. However, all neighbouring simplices can
still satisfy the zero net spin constraint. One therefore expects an infinitesimal
magnetic field, say ${\mathbf h} = \epsilon \hat{z}$ to immediately
polarize the net spin of the triangle with two defects, while having
no effect on any other simplex.

What is the total spin of the resulting state? Since each physical spin
is shared by two simplices, and only one simplex has non-zero net spin, we may write $S^z_{\mathrm{tot}} = \frac{1}{2}
\left (\sum_{\XBox}S^z_{\XBox} + \sum_{\triangle} S^z_\triangle \right) = 3/4$!
Thus, such a correlated defect gives rise to a spin of $S=3/4$---these
have been refered to in the literature as `half-orphans'~\cite{Henley_2000},
and consitute the second of our promised surprises.

To complete our discussion, we must also note that these half-oprhans
come with a statutory warning: As in our earlier example, this is again a purely $T=0$ statement relying
on minimizing the interaction energy. Again, it is not at all obvious
that any of this survives the entropic effects of thermal fluctuations
at non-zero temperature. Indeed, unlike in the previous example in which
entropic effects have been analyzed and are now well-understood, not much
is known about the effects of thermal fluctuations on these half-orphans, although
diluted SCGO has been studied using computer simulations
and phenomenological approaches~\cite{Schiffer_daruka,Moessner_berlinsky}. It thus remains
an open question whether these $S=3/4$ spins survive the effects of
non-zero temperature and can be `seen' in experiments,
and we are working on providing some definite answers soon~\cite{Sen_damle_moessner_ongoing}.

{\em Acknowledgements: } The author acknowledges useful discussions with
D.~Dhar, and with his collaborators A.~Banerjee, D.~Heidarian, R.~Moessner,
and A.~Sen. Funding from DST SR/S2/RJN-25/2006, and the hospitality of School of Physical Sciences,
JNU and Physics Department, IIT Bombay during the completion of this
contribution is also gratefully acknowledged.

\end{document}